# ELECTRONIC CONTROL OF EDGE-MODE SPECTRUM OF INTEGER-HALL-EFFECT 2D ELECTRON WAVEGUIDES

## GUENNADI A. KOUZAEV

Department of Electronics and Telecommunication, Norwegian University of Science and Technology-NTNU, Trondheim, No-7491, Norway, E-mail: guennadi.kouzaev@iet.ntnu.no

**Abstract.** In this paper, the control of the edge-mode spectrum of integer-Hall-effect 2D waveguides by the electric field is proposed and modeled with the effective mass approach. Under certain found conditions, the applied transversal electric field allows refining the modal spectrum from non-localized waves, and, additionally, it can switch the edge-mode from the propagating to the evanescent state, and it is interesting in the design of the edge-mode off/on logic components. These waveguides, arbitrary biased by potentials, are described by the Pauli spin-less charge equation, and they are simulated using the order reduction method of partial differential equations in its Kron's quantum-circuit representation. Additionally to the spectrum control mechanism, the influence of large-scale disorder of confinement potential and magnetic field on the edge localization and modal switching is studied.

## 1. Introduction

The goal of this paper is to discover the electronically-controlled mechanism of the edge-mode formation in the integer-Hall-effect 2D waveguides and their off/on switching by the electric field in the conditions perturbed by the randomness of potentials.

In the general case, the behavior of an electron in external electromagnetic (EM) field is described by the quantum-mechanical equation proposed by W. Pauli in 1927 [1]. A particular case of this equation for a spin-less charge in open space was analytically solved by L. Landau who found the magnetic fields allowing solvable equations and discretization of electron energy levels in the magnetic field [2]. This effect causes quantization of the Hall-transversal resistance of 2D electron waveguides [3], which was discovered in 1980, and it is known today as the *integer quantum* Hall effect because of this resistance is proportional to an integer number of a quantum level [4]. The integer Hall effect can be explained or modeled using calculations of single non-interacting electrons or quasi-single electrons in 2D waveguides using the effective mass method [5,6].

Besides the integer Hall-effect, the *fractional* one is known to appear in 2D/3D multi-electron waveguides [8-11]. In these quantum conductors, the charge carriers interact with each other and with heavy molecules, and many interesting phenomena are known including the edge modes appearing with or without external magnetic field and at low or room temperatures [12-14]. However, according to common opinion [5,11,12], the electron-electron interactions only disturb the underlying physics of Hall Effect.

Formation of these modes is influenced by parasitic randomness of electric and confinement potentials, and these effects are studied, for instance, in [5,11,15]. In general, these weak disorders can be suppressed by higher-level magnetic fields which level should be defined during design work.

In general, the topologically protected edge modes are interesting in noise-immune edge-mode signaling and in electronic logic, and they may find their place in quantum computing.

Although many results in this field are known, some problems still hinder the applications. For instance, the modal spectrum of topological insulators is contaminated by non-localized modes, and a technique on filtering of modes is required to be developed. Electronic control of these modes will allow designing new logic components [16] and further research is required in this direction.

Taking into account that the quantum Hall-effect components can be interesting in practice, there is a strong need to study in more details the background effects, including the purification of modal spectrum from non-localized waves, the edge-mode switching by electric field, and modeling the influence of occurring-in-practice randomness of potentials on the modal formation and their switching.

For the sake of simplicity, a 2D integer Hall-effect waveguide is chosen as the study object because of the fundamentality of its modeled properties. Following [5,10-13] it is supposed that these found properties may be common as well to more complicated in physics fractional Hall-effect waves. For simulation, the effective mass approach is used earlier applied to the spin-waveguides by many authors.

In Section 2, the waveguide geometries are given and considered. To calculate waveguides, a known algorithm of representation of Schrödinger equation by two partial differential ones of the first order is used which allows the circuit representation familiar to practicing microelectronic engineers (Section 3).

Introductory material on the constant-confinement potential waveguides biased by the magnetic field is considered in Section 4. The results on modeling of the integer Hall effect in periodical and pseudo-random potential nano-engineered waveguides are in Section 5. The influence of the randomness of the biasing magnetic field is studied in Section 6. In Section 7, the proposed technique of joint electronic and magnetic purification of the waveguide's spectrum from the non-localized modes is given, and the electronic switching of the effect of edge modes is discovered. The conclusions are in Section 8.

## 2. The geometry of Studied Electron Waveguides Biased by Magnetic and Electric Fields

A two-dimensional electron waveguide, in which the waves propagate ballistically along the $z-$axis, is shown in Fig. 1a. Particularly, it can be formed in vertically ($y-$direction) grown sandwiches of $Ga_{1-x}Al_xAs$ - GaAs - $Ga_{1-x}Al_xAs$ with the quasi-2D channel in gallium arsenide [5,6,17]. As well, more complicated heterostructures can be used to form the 2-D electron waveguides.

One of the widely used approaches to describe these waveguides is the effective-mass method that takes into account the interaction of electrons with the semiconductor. In this case, quantum-mechanical many-body physics is described by a single-particle Schrödinger equation regarding the *envelope wavefunction* [5,17-19] (below just wavefunction) $\psi(x,z)$ and biased by confinement potential $U(x)$. This method is widely used in quantum electronics and it is matched well with the experimental and analytical results (see, for instance, [17,20-22]), although some restrictions are with the very thin layers and sub-nanometric inclusions [17]. This approach is applicable to the electron waveguides biased by electric and magnetic fields [5,6,17] as well.

The considered in this paper waveguide is of the limited width $(0,a)$ and the ideal hard-walls $\left(\psi(x=0)=\psi(x=a)=0\right)$ caused by the infinite values of confinement potential at $x=(0,a)$ (Fig. 1).

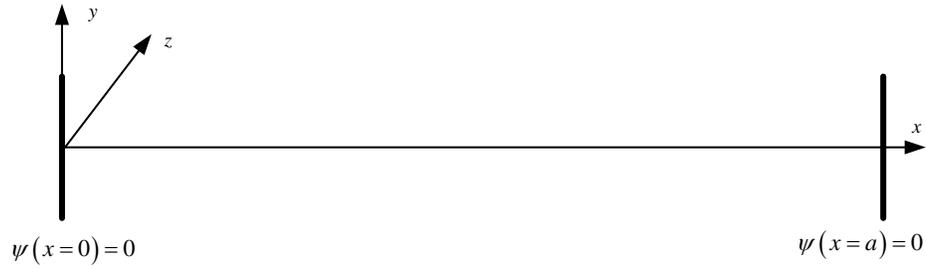

a

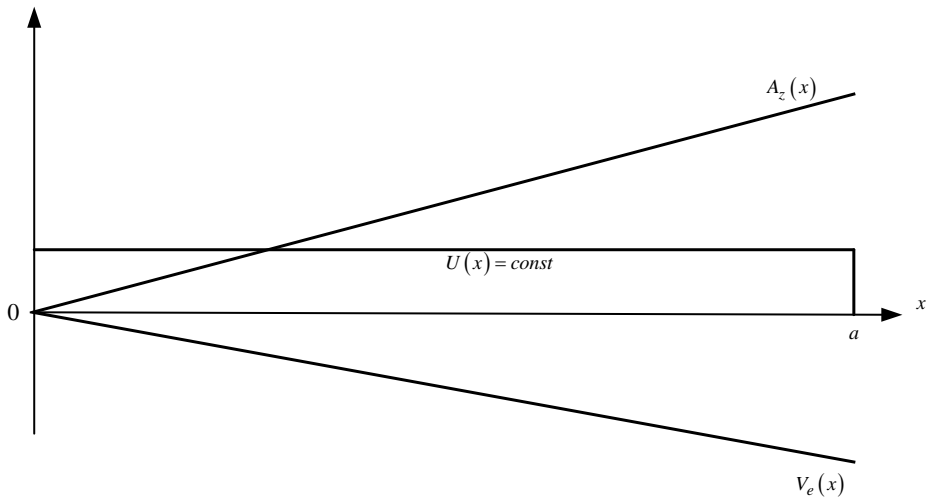

b

Fig.1. 2D electron waveguide geometry (a) and the potential function distributions $\left( U,\ V_e,\ A_z \right)$ along the waveguide cross-section (b).

The forward electron modes are propagating along the $z-$axis, and the backward ones are against it, which is normal to the picture plane in Fig. 1.

Particularly, this waveguide can be biased with longitudinal vector magnetic potential $A_z = B_0\,y$, where $B_0$ as the magnetic field magnitude and with scalar electric potential $V_e\left(x\right) = -e\mathrm{E}_0^{(x)}x$, ( $e = 1.620218\cdot10^{-19}$ C ) is the electron charge and $\mathrm{E}_0^{(x)}$ is the intensity magnitude of the $x-$oriented electric field), and the confinement potential $U\left(x\right)$ which is in this paper of the constant (Fig. 1b), regular-lattice, and random-lattice shapes.

The non-uniformity of this potential $U\left(x\right)$ is with the lateral modulation of two-dimensional engineering technology [15-17,23-25], or it can be caused by random impurities of waveguides along the $x-$axis [5].

### 3. Spin-less Charge Schrödinger Equation and its Treatment by Equivalent Circuit Approach

Spin-less charge in a magnetic field is described here with the reduced Schrödinger-Pauli time-independent equation (1) that does not take into account the electron spin and written with the constant effective mass $m_{\text{eff}}$ approximation [2,5,6,17]:

$$\left[\frac{1}{2m_{\text{eff}}}\left(\frac{\hbar}{i}\nabla - e\mathbf{A}(x)\right)^2 + U(x) + V_e(x) - E\right]\psi(x,z) = 0 \tag{1}$$

where $\hbar = 1.05457 \cdot 10^{-34}$ J$\cdot$s is the normalized Plank constant, $\nabla \equiv \mathbf{x}_0\frac{\partial}{\partial x} + \mathbf{z}_0\frac{\partial}{\partial z}$, $\mathbf{x}_0$ and $\mathbf{z}_0$ are the unit vectors, $i$ is the imaginary unit, $\mathbf{A}(x)$ is the magnetic vector potential, $E$ is the total electron energy. The effective mass of electron $m_{\text{eff}}$, in general, depends on the $x$-variable, and it should be calculated quantum-mechanically. In more accurate models, the effective mass is anisotropic as well. For the sake of simplicity [5,6,17], let's take $m_{\text{eff}} = const = 0.067m$, $\left(m = 9.109 \cdot 10^{-31} \text{ kg}\right)$, as it is for GaAs material.

Consider this equation as a mean to calculate the modal longitudinal propagation constants $k_z$ as the eigenvalues of Equ. (1). Supposing that the operator $\left(\frac{\hbar}{i}\nabla - e\mathbf{A}(x)\right)$ is commutative and following to the Landau's guage rule [2], the magnetic vector potential is chosen, particularly as $\mathbf{A} = \mathbf{z}_0 B_0 x$ although these equations can be solved with other directions of the magnetic field [6]. Then $\mathbf{B} = -\nabla \times \mathbf{A}$ has only one $B_y = B_0$ component oriented along the $y-$axis.

Consider simulating only the backward propagating modes traveling against the $z-$axis, $\psi(x,z) = \psi(x)e^{-ik_z z}$ ($k_z$ is the modal propagation constant).

Following the Landau's gauge rule (see above) and the mentioned wavefunction longitudinal dependence, Equ. (1) is transformed into

$$\left[-\frac{\hbar^2}{2m_{\text{eff}}}\frac{d^2}{dx^2} + V(x) - E\right]\psi(x) = 0 \tag{2}$$

where (see as well [2,5,17])

$$V(x) = \frac{e^2 B_0^2 x^2}{2m_{\text{eff}}} + \frac{\hbar e B_0 k_z x}{2m_{\text{eff}}} + \frac{\hbar^2 k_z^2}{2m_{\text{eff}}} - e\mathrm{E}_0^{(x)}x + U(x). \tag{3}$$

As it is seen from (3), $V(x) \sim \mp k_z$ and it defines the direction-depending properties of matter waves. Additionally, the waves depend on the orientation of the magnetic and electric fields because $V(x) \sim \pm B_0$ and $V(x) \sim \pm \mathrm{E}_0^{(x)}$.

An analytical solution of Equ. (2) is considered in [2,5] for an infinite plane with $U(x) = 0$. The wavefunctions are found as the Hermite polynomials multiplied by exponents. A numerical model for a 2D electron waveguide of limited width and zero potential and the electric field is given in [5,7].

Below, a more general theory is used for calculation of an electron waveguide controlled by magnetic, electric and confinement potentials. This theory is based on the method of the order reduction of the Schrödinger equation proposed in [26] and further elaborated and verified experimentally for many applications [21,22, 27-36].

Following these contributions, Equ. (2) is transformed into a system of the first-order differential equations regarding the equivalent quantum "voltage" $u(x) = \psi(x)$ and Kron's[1] "current" $j(x) = i\frac{\hbar}{m_{\text{eff}}}\frac{du(x)}{dx}$ for general $V(x)$, i.e. the method of the reduction of the order of the differential equations is used :

$$\frac{du(x)}{dx} = -i\frac{m_{\text{eff}}}{\hbar}j(x),$$
$$\frac{dj(x)}{dx} = \frac{2i}{\hbar}(V(x) - E)u(x).$$
(4)

Implementing equ. (4) with the condition $u(x = 0, a) = 0$, we obtain a boundary value problem, which solution gives us the propagation constant $k_z$ and modal wavefunction $\psi(x)e^{-ik_z z}$ for any backward-propagating ($z-$ axis) modes while the electron energy $E$ is fixed. Similarly, the forward-propagating ($z-$ axis) modes are treated.

Consider the $x-$ dependence, particularly, Equ. (4) is the Kirchhoff's law for the voltage $u(x)$ and current $j(x)$ at any point of this axis. Suppose that it is divided into a number $K$ of domains $\Delta x_k$ where the potential $V(x_k) = const$ (Fig. 2). Then, these domains can be considered as the short lengths of a transversal transmission line, which have their propagation constants $k_x^{(k)}$ and characteristic impedances $Z_c^{(k)}$ and described by their transfer matrices $A^{(k)}$ (see for more details [36]):

$$A^{(k)} = \begin{pmatrix} \cos k_x^{(k)}\Delta x_k & iZ_c^{(k)}\sin k_x^{(k)}\Delta x_k \\ \dfrac{i}{Z_c^{(k)}}\sin k_x^{(k)}\Delta x_k & \cos k_x^{(k)}\Delta x_k \end{pmatrix}.$$
(5)

Here, $k_x^{(k)} = \sqrt{\frac{2m_{\text{eff}}}{\hbar^2}(V(x_k) - E)}$, and $Z_c^{(k)} = \frac{\sqrt{m_{\text{eff}}}}{\hbar k_x^{(k)}}$ .

Next step in this algorithm is choosing an arbitrary reference point $X_0$ (Fig. 2) at which the left $\overleftarrow{Z}$ and right $\overrightarrow{Z}$ impedances are obtained using the multiplication of corresponding matrices $A_k$ .

---

[1] This Kron's current is different from the quantum-mechanical current defined as $j(x) = i\hbar/2m_{\text{eff}}\left(\psi^*\,d\psi/dx - \psi\,d\psi^*/dx\right)$

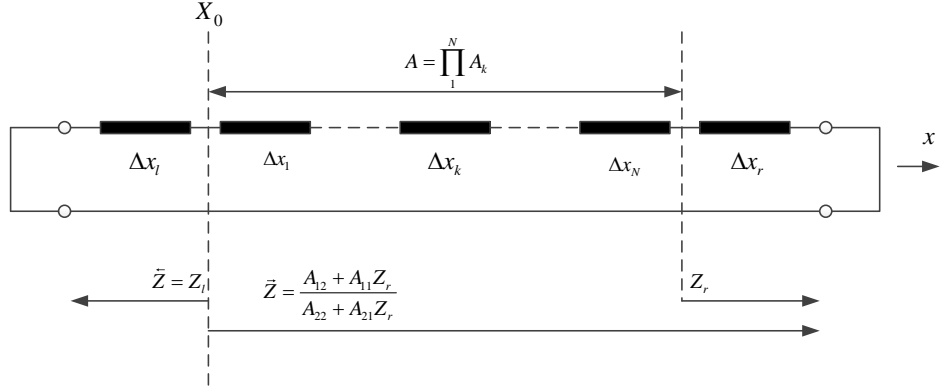

Fig. 2. Transversal transmission line model for 2D electron waveguide. The left- and right-side shortenings correspond to the hard walls of this waveguide at $x = 0, a$, correspondingly.

To find the longitudinal propagation constant $k_z$, the condition of the transversal resonance is written at $x = X_0$ (Fig. 2):

$$\breve{Z} + \vec{Z} = 0 \quad . \tag{6}$$

It gives a nonlinear and transcendental equation regarding $k_z$. To calculate the voltage and current distributions along the $x-$axis, its source is placed at the reference point $X_0$ and their amplitudes are obtained using an iteration scheme [37]:

$$\begin{pmatrix} u_k \\ j_k \end{pmatrix} = A_k \begin{pmatrix} u_{k+1} \\ j_{k+1} \end{pmatrix} \quad . \tag{7}$$

This algorithm requires inversion of $2 \times 2$ matrices $A_k$, which is performed here analytically. Proper normalization [36] of these matrices allows avoiding or essentially decrease the instability, which is typical for most calculations with the transfer matrices [38]. Additionally, the voltage or wavefunction is normalized as

$$\int_0^a u(x) \frac{1}{m_{\text{eff}}} u^*(x) dx = 1 \tag{8}$$

where the star sign $(*)$ means the conjugation operation over complex, in general, $u(x) = \psi(x)$.

## 4. Edge-mode Formation in Waveguides of Constant Potential

Although this case has been studied in several contributions, our modeling results are included in this paper as the basic ones needed for introduction, testing and a better understanding of physics and the used algorithm.

Our simulations start with a waveguide whose confinement potential $U(x) = const$, while the magnetic field is changed discretely from domain to domain. In Fig. 3, this potential was chosen at the level $U(x)/E = U_0/E = 0.8667$ with electron energy $E = 0.1$ eV. The overall number of domains is held $K = 1060$ that supports practically complete convergence of results in most calculations performed for the waveguide's width $a = 58.5$ nm [36].

For the magnetic field $B_0 = 0$, the obtained propagation constant $k_z$ was compared with the analytically calculated value $k_z = \sqrt{\dfrac{2m_{\text{eff}}}{\hbar^2}\left(E - U_0\right) - \left(\dfrac{\pi}{a}\right)^2}$, and no difference was found for six meaningful digits after the period sign. The wavefunction is found the same with the corresponding function of the mentioned homogeneous 2D electron waveguide calculated analytically (Fig. 3, solid-blue line).

To study the effect of magnetic field, it increases gradually, and a set of calculations of propagation constant and squared envelope wavefunction $|\psi|^2$ of only main mode was performed here because for the given waveguide geometry, particle energy, and potential, the higher-order modes are evanescent, i.e., they are not propagating. The propagation constant dependence was calculated for $B_0 = 0 - 5.12$ T, and it tends to zero with this magnetic field.

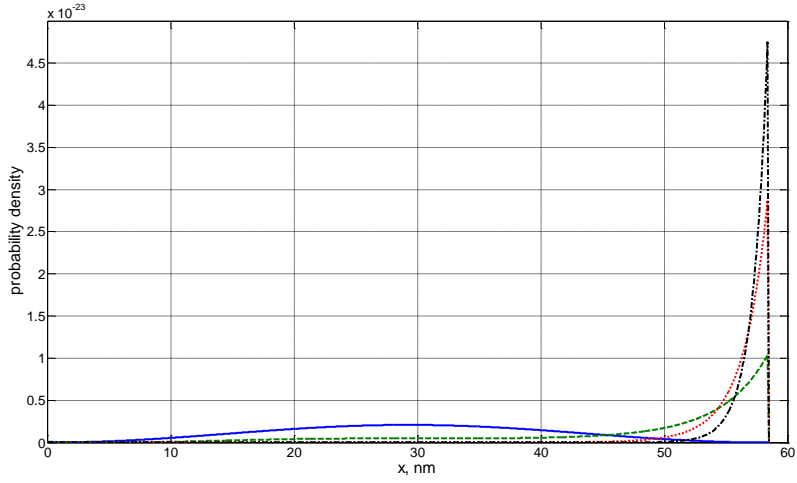

Fig. 3. Probability-density shapes in regular potential electron waveguide. $U_0/E = 0.8667$ and $a = 58.5$ nm. Blue-solid line: $B_0 = 0$ T; green-dash line: $B_0 = 1.5$ T; red-dot line: $B_0 = 3$ T; black dot-dash line: $B_0 = 5.12$ T.

Wave-function shape evolution starts with $\sim \sin^2\left(\dfrac{\pi x}{a}\right)$ for $B_0 = 0$ (solid-blue curve). According to theory of Hall-effect in 2D electron waveguides, the main mode of this waveguide is localized with $B_0$, and it is transformed into an edge-type mode, which pick is "pressed" to the right side of waveguide due to the wave's backward type. Due to the properly chosen geometry, potential, and electron energy this waveguide is monomodal, and the electron transportation by this edge mode can be realized, in theory, without contamination of this process by non-localized waves which was found harmful in many cases.

## 5. Edge-mode Formation in Waveguides of Periodical and Quasi-random Lattice Confinement Potential

The transversal (along with the $x-$axis) potential modulation can be caused technologically using different methods of lateral potential engineering [15-17,23-25]. Particularly, the modulated potentials can be close to being periodical, and the imperfection of technology may randomize them. These cases are calculated being again with the above-described model (1)-(8) that can be improved using more complicated effective mass algorithms.

The transversally-modulated electron waveguides are studied here starting the simulation of the *regular-lattice potential* $U_{\text{lattice}}(x)$ waveguides as the reference structures (Figs. 1b, 4). This potential is generated according to the following formula [36]:

$$U_{\text{lattice}}(x) = U_0 \cdot \text{rem}(N, 2) \qquad (9)$$

where $U_0$ is the magnitude coefficient, and $rem(2, N)$ is the Matlab reminder-after-division function.

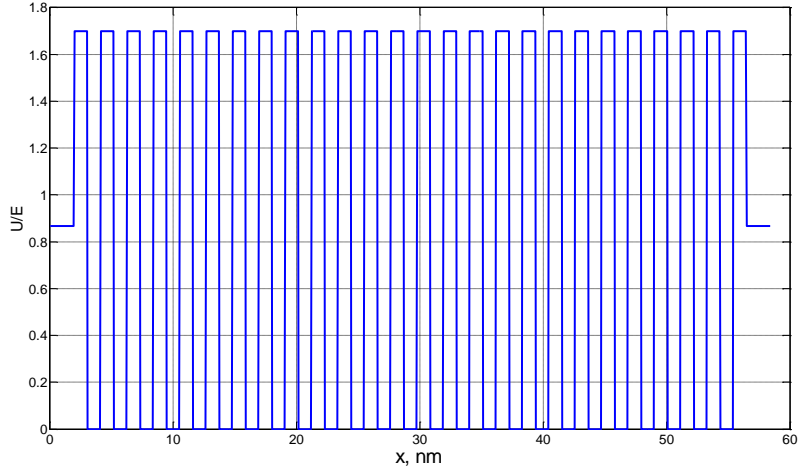

Fig. 4. Regular-lattice potential $U_{\text{lattice}}(x)$ calculated according to (9) with $U_0 / E = 1.7$ given for a 2D-waveguide. Other waveguide parameters: $a = 58.5$ nm and $N = 51$. Additional left and right potential steps given for $\Delta x_l = 2$ nm and $\Delta x_r = 2$ nm (see Fig. 2) are on the level $0.5 U_0$.

With the magnetic-field increase from $B_0 = 0$ to 6 T, the longitudinal propagation constant $k_z$ decreases similarly to the waveguide with a constant potential.

A set of probability-density plots corresponding to these simulations is in Fig. 5.

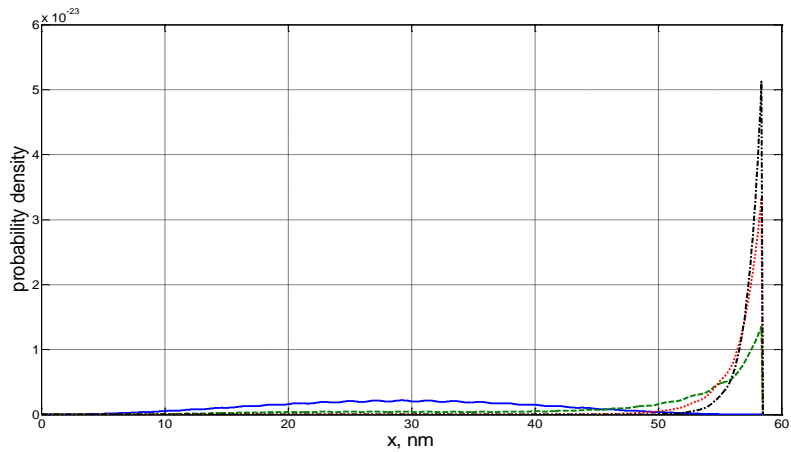

Fig. 5. Probability-density shapes in regular lattice potential electron waveguide. $U_0 / E = 1.7$, $a = 58.5$ nm, and $N = 51$. Blue-solid line: $B_0 = 0$ T; green-dash line: $B_0 = 1.5$ T; red-dot line: $B_0 = 3$ T; black dot-dash line: $B_0 = 5$ T.

It is seen that this lattice-shaped potential $(U_0/E = 1.7)$ prevents the edge-mode formation. It occurs at a stronger magnetic field $B_0$ in comparison with the constant-potential waveguide (Figs. 1a, 3). An additional effect of this lattice is that this waveguide starts to be multimodal for the same geometry and electron energy as in the case of Fig. 1a due to lower averaged potential.

The regular lattice potential can be perturbed by technological imperfections, and the potential steps can be random. These *pseudo-random lattice potential waveguides* can form the localized modes of the Anderson type if their lattice step $\Delta x_k$ is comparable with $\lambda/2\pi = \hbar/\sqrt{2m_{\mathrm{eff}}E}$ and the number of needed steps is overall several hundred that was found in many optical, microwave and acoustic waveguides which media is randomly perturbed [39,40]. In general, this strong localization may prevent the formation of edge modes caused by the Hall effect. In narrow waveguides with a smaller number of these large potential steps as in the considered one, only weak localization was found [36].

Following [36], the pseudo-random potential $U_{\mathrm{prandom}}(x)$ (Fig. 6) is generated as

$$U_{\mathrm{prandom}}(x) = U_0 \cdot \mathrm{rem}(N, 2) A(N) \text{ where } A(N) \text{ is } N- \text{ sample pseudo-random data obtained using the Matlab's}$$

*rand.m* function in its default regime in this case.

(10)

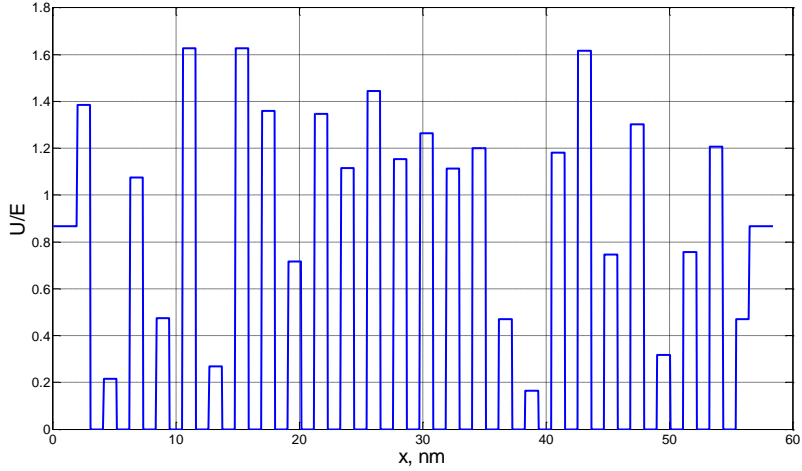

Fig. 6. Pseudo-random lattice potential $U_{\mathrm{prandom}}(x)$ according to (10) with $U_0/E = 1.7$. Waveguide parameters: $a = 58.5$ nm and $N = 51$. The additional left and right potential steps for $\Delta x_l = 2$ nm and $\Delta x_r = 2$ nm (see Fig. 2) are chosen on the level of the averaged value of this pseudo-random potential.

In [36], the formation of localized modes is studied in pseudo-random lattice potential, and it is noticed on the combined mechanism formation of localized modes in these waveguides. The randomness and reflections from the ideal "hard" walls form some localized shapes, which are found relative stable towards a variation of random potential sets.

One of the modal shapes, corresponding to the potential shown in Fig. 6, is in Fig. 7 (blue solid line). It is formed at the right waveguide's side if $B_0 = 0$. The extreme of this curve may be connected with the local lacunas of this random potential (Fig. 6) and they are rather unstable with the variation of this potential, as it was shown in [36]. This localization is easily destroyed with the increase of the magnetic field up to 6 T, and the sharp edge-mode shape is after $B_0 > 2 - 2.5$ T (red-dash and black dash-dot lines). As in the above-described cases, the modal propagation constant decreases monotonically with the magnetic field. One of the findings in [36], is localized by pseudo-random potential main modes which sharp picks are shifted to the one side of the waveguide. They are formed, mostly, in the waveguides with large-width potential steps $\Delta x_k > h / \sqrt{2 m_{\text{eff}} E}$ (Fig. 8).

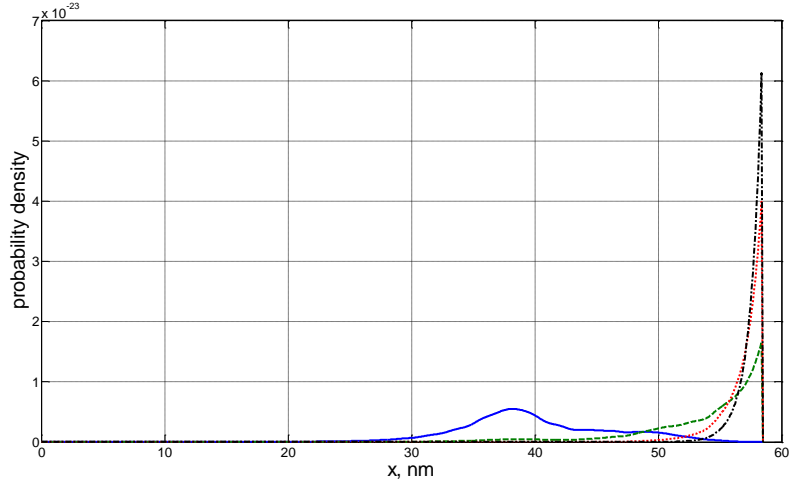

Fig. 7. Probability-density shapes in pseudo-random lattice potential (Fig. 7) waveguide. $U_0 / E = 1.7$ and $a = 58.5$ nm. Blue-solid line: $B_0 = 0$ T ; green-dash line: $B_0 = 1.5$ T ; red-dot line: $B_0 = 3$ T ; black dot-dash line: $B_0 = 5$ T .

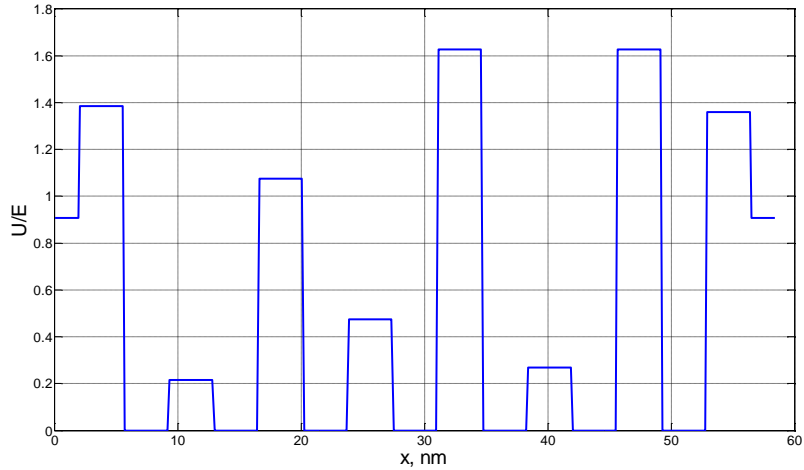

Fig. 8. Pseudo-random lattice potential $U_{\text{prandom}}(x)$ calculated according to (10) with $U_0 / E = 1.7$ . $a = 58.5$ nm , and $N = 15$ . The additional left and right potential steps given for $\Delta x_l = 2$ nm and $\Delta x_r = 2$ nm (see Fig. 2) are chosen on the level of the averaged value of this pseudo-random potential.

These modal shapes (Fig. 9, blue solid line) found to be rather stable being the products of interference of reflected partial transversal waves from the ideal hard walls at $x = 0, a$ and from the random-potential steps.

It is seen from this picture, that the non-biased by magnetic field pseudo-random potential waveguides can deliver the same shapes (Fig. 9, blue solid line) as those in the magnetic field (Fig. 9, black-dash line) being different in physical origin.

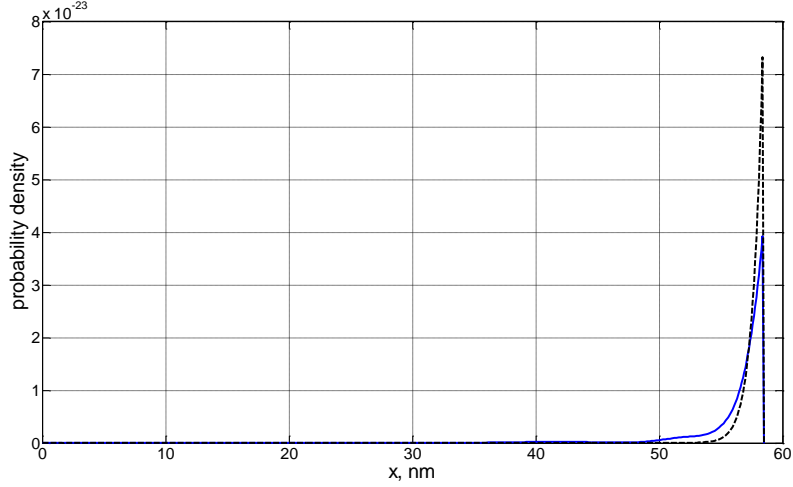

Fig. 9. Probability-density shapes in pseudo-random lattice potential (Fig. 8) electron waveguide. $U_0/E = 1.7$, $\Delta x_l = \Delta x_r = 2$ nm (see Fig. 2), and $a = 58.5$ nm. Blue-solid line: $B_0 = 0$ T and the black-dash line: $B_0 = 5$ T.

It is interesting to research how these two mechanisms of formation of localized shapes are working together. In this purpose, it has been performed a statistical study of the stability of the pick shapes taking into account the random content of the potential $V(x)$. We generated 15 sets of pseudo-random potential with $U_0/E = 1.7$ and $N = 15$ using the default and shuffle regimes of the Matlab's *rand.m* function. The left and right potentials from Fig. 8 are kept on the averaged level of the generated lattice.

For each of the three values of the magnetic field $(B_0 = 0, 1, 3 \text{ T})$, we performed 15 calculations of the wavefunction. It was found that the stability of the mentioned pick shape was growing with the magnetic field. For instance, at $B_0 = 0$ and 1 T, only 8 shots (from 15) preserved the sharp pick. At $B = 3$ T, already 14 shots (from 15) gave the localized wavefunctions. Here, the randomness of confinement potential $U(x)$ plays a perturbative role, and these picks can be even sharper with the combination of pseudorandom potential and magnetic field instead of separate biasing by these fields.

### 6. Magnetic Field Randomness Influence on the Edge-Mode Formation

It is interesting to study the influence of *randomness of the magnetic field*. It may occur due to magnetic clusters of molecules randomly distributed in some materials [16]. There, it was found experimentally that the edge-mode formation due to the fractional spin quantum Hall effect was destroyed in many cases.

Although we model the integer quantum Hall effect, this may take place in our case, too. Unfortunately, Equ. (1) is handled further here only if the magnetic field is constant in the $x-$direction because of the requirements to satisfy the Landau gauge condition. To keep this requirement, we suppose that the magnetic field is constant in the limits of each domain $\Delta x_k$ only. But the value of the magnetic field $B_0^{(k)}$ can be varied randomly from $k-th$ domain to another (Fig. 10).

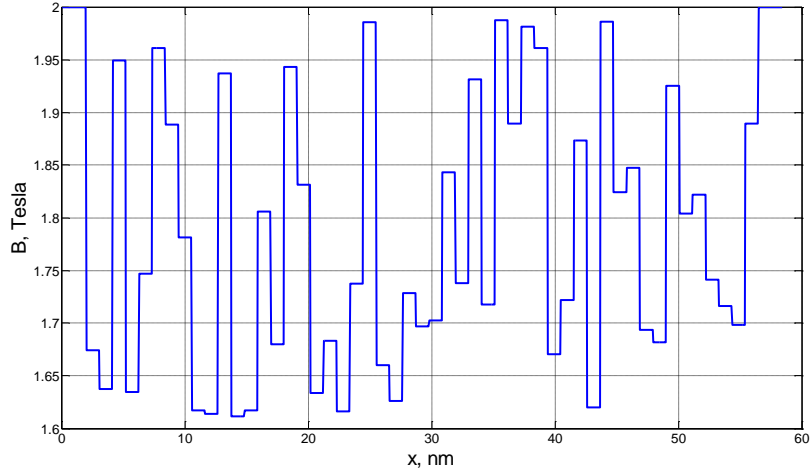

Fig. 10. Pseudo-random magnetic field magnitude $B_0^{(k)}$ generated according to (11). $\alpha = 0.2$, $B_0 = 2$ T, $a = 58.5$ nm, $\Delta x_l = \Delta x_r = 2$ nm, and $N = 51$.

It is calculated according to the following formula

$$B_0^{(k)} = B_0(1 - \alpha A(k)) \qquad (11)$$

where $A(k)$ is from the pseudo-random set $A(N)$ generated using the Matlab's *rand.m* function in its default state, and the parameter $0 < \alpha < 1$ defines the randomness magnitude. Some simulation results are shown in Fig. 11.

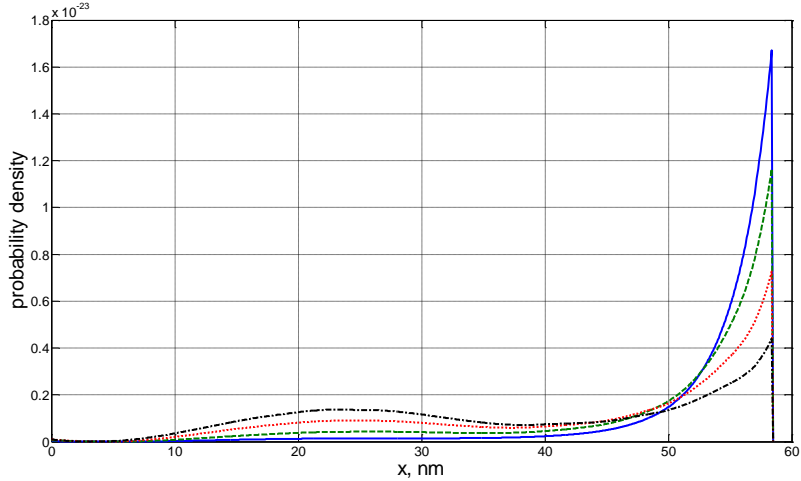

Fig. 11. Probability-density shapes in electron waveguide biased by the pseudo-random magnetic field (Fig. 12) and constant potential $U_0/E = 0.8867$. Blue-solid line: $\alpha = 0.1$, green-dash line: $\alpha = 0.5$; red-dot line: $\alpha = 0.8$; black dot-dash line: $\alpha = 1$. $B_0 = 2$ T, $a = 58.5$ nm, and $N = 51$.

It is seen that the essential delocalization of wavefunction starts with which is following the experiments [16] on the harmful influence of random magnetic clusters on the fractional spin quantum Hall effect.

## 7.    Control of Edge Modes with Electric Field

In general, the Hall-effect waveguides are very polluted by non-localized waves which makes hard using only the edge modes for electron transportation. Great interest is in suppressing these non-localized waves and in electronic switching devices for edge modes [41,42-44]. Here, some analogies with the EM waves can be used.  Many results are in microwaves where the modes are filtered or switched according to their spatial topologically-stable content using passive and semiconductor components [35].

Consider Fig. 3, where the probability density in a monomodal waveguide is shown. Unfortunately, this one-mode transportation has its own drawback because of its poor signal-to-noise ratio, and the signaling by multiple edge modes is preferable [41]. Additional freedom in material and control of its confinement potential allows partly to solve this problem. Choosing an appropriate combination of electron energy, potential, and geometry of waveguide, it is made multimodal.  Fig. 12 shows the propagation constant $k_z$  dependence on biasing magnetic field $B_0$  for the found three propagating modes.

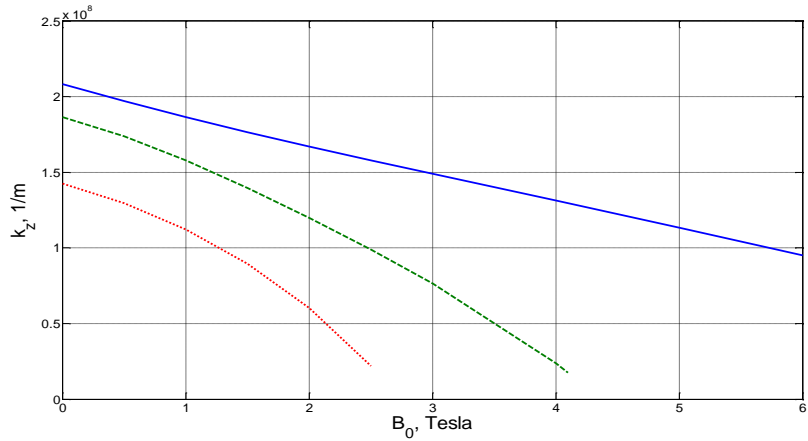

Fig. 12. Propagation constants $k_z$  of the first three modes versus magnetic field  $B_0$ .  $U_0/E = 0.7367$  and $a = 58.5$ nm . Main mode: blue-solid line; 2nd mode: green-dash line; 3rd mode: red-dot line.

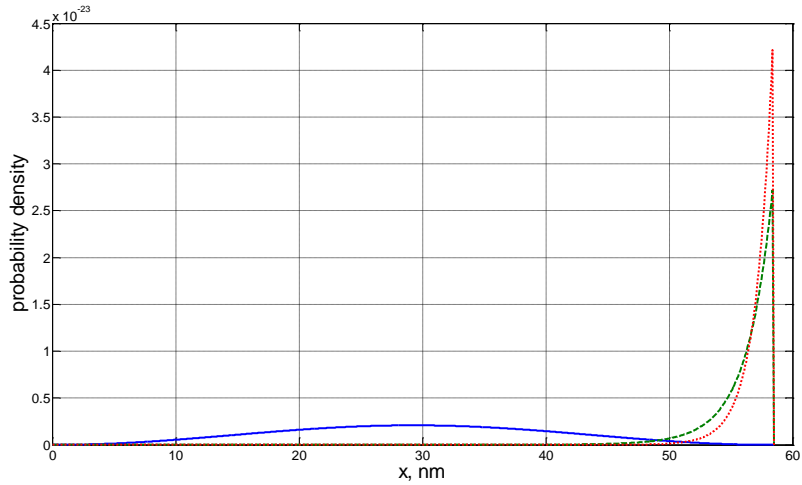

Fig. 13. Probability-density shapes of the main mode in constant-potential. Blue-solid line: $U_0/E = 0.7367$ and $a = 58.5$ nm , $B_0 = 0$ T ; green-dash line: $B_0 = 2.5$ T ; red-dot line: $B_0 = 4$ T .

The mentioned potential is of 85% of that for the monomodal waveguide (Figs. 3) in this case. Simulation of wavefunctions with the magnetic-field increase shows that localization starts with the main mode (Fig. 13) and it is following with the second one (Fig. 14).

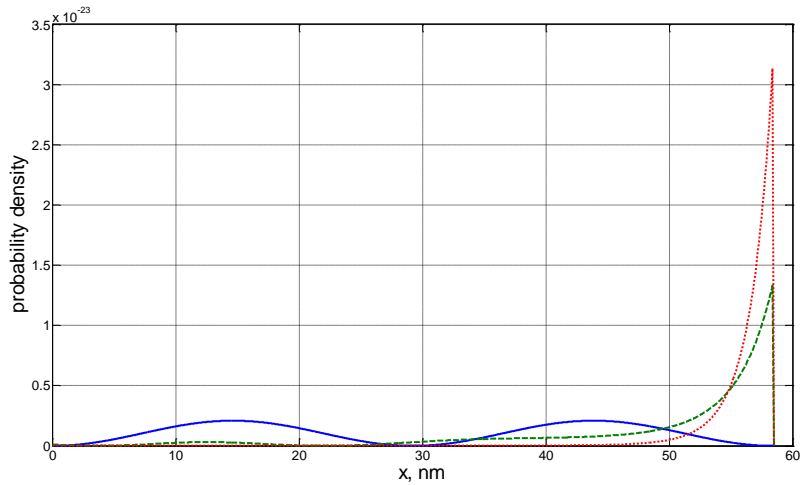

Fig. 14. Probability-density shapes of the 2$^{nd}$ mode in constant-potential. $U_0/E = 0.7367$ and $a = 58.5$ nm . Blue-solid line: $B_0 = 0$ T ; green-dash line: $B_0 = 2.5$ T ; red-dot line: $B_0 = 4$ T .

The third mode is coming to be non-propagating in its non-localized state in this waveguide at $B_0 = 5.2$ T (Fig. 15).

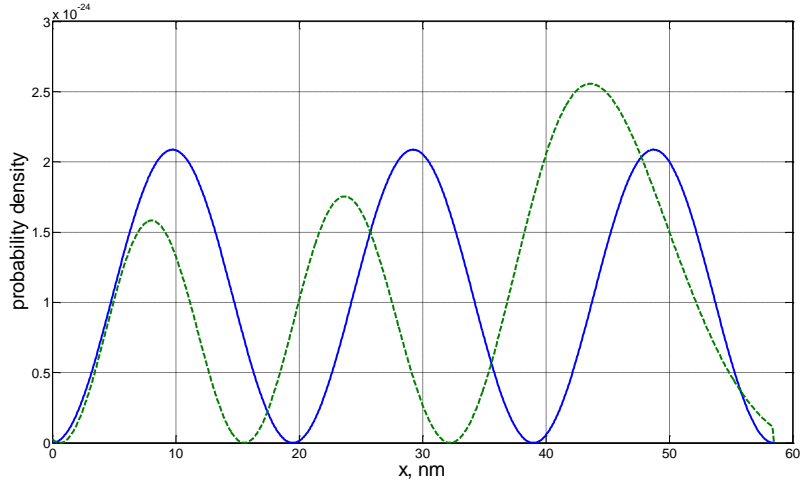

Fig. 15. Probability-density shapes of the 3$^{rd}$ mode in constant-potential electron waveguide. $U_0/E = 0.7367$ and $a = 58.5$ nm . Blue-solid line: $B_0 = 0$ T ; green-dash line: $B_0 = 2.5$ T .

Then, for the magnetic field over $B_0 \geq 5.2$ T , *electron* transportation can be realized by only two edge-localized modes.

Although there are several mechanisms on how to vary the confinement potential electronically, the edge-mode transportation can be controlled by the applied in the $x-$direction electric field without changing the potential $U(x)$ (see equ. (1)).

For instance, Fig. 16 shows the evolution of wavefunction with the electric field $\mathrm{E}_0^{(x)}$ increase.

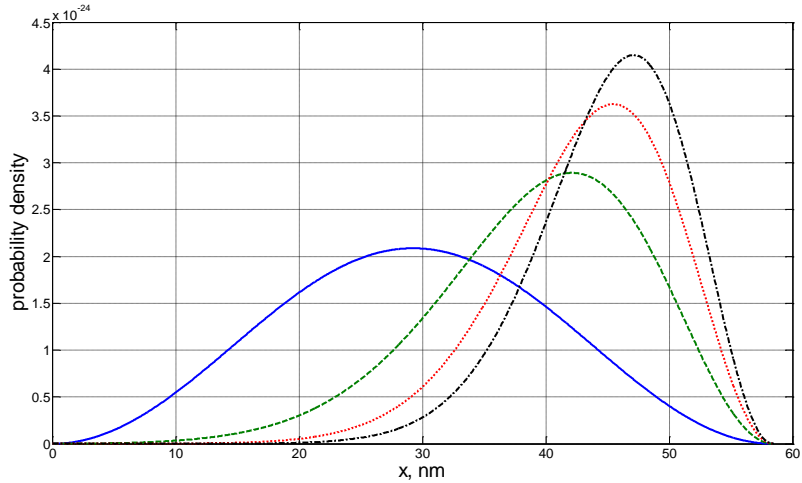

Fig. 16. Probability-density shapes in electron waveguide biased by electric field $\mathrm{E}_0^{(x)}$ . $B_0 = 0$ T , $U_0/E = 0.8667$ , and $a = 58.5$ nm . Blue-solid line: $\mathrm{E}_0^{(x)} = 0$ V/m ; green-dash line: $\mathrm{E}_0^{(x)} = 3 \cdot 10^5$ V/m ; red-dot line: $\mathrm{E}_0^{(x)} = 6 \cdot 10^5$ V/m ; black dot-dash line: $\mathrm{E}_0^{(x)} = 9 \cdot 10^5$ V/m .

It is seen that electron tends to the side with lower potential (see equ. (3)), and the wavefunction is localized slightly at the right part of the waveguide. The propagation constant increases with the electric field, and it can be used for electronic control of propagation characteristics of modes.

Applying a strong electric field to waveguides biased by weak magnetic field leads to the destruction of edge localization caused by opposite signs of terms with the magnetic and electric fields in (3), and it is seen from Fig. 17, *where electric field transforms localized shape (solid line) into a main-mode shape deformed by the electric field (dot-dash line).*

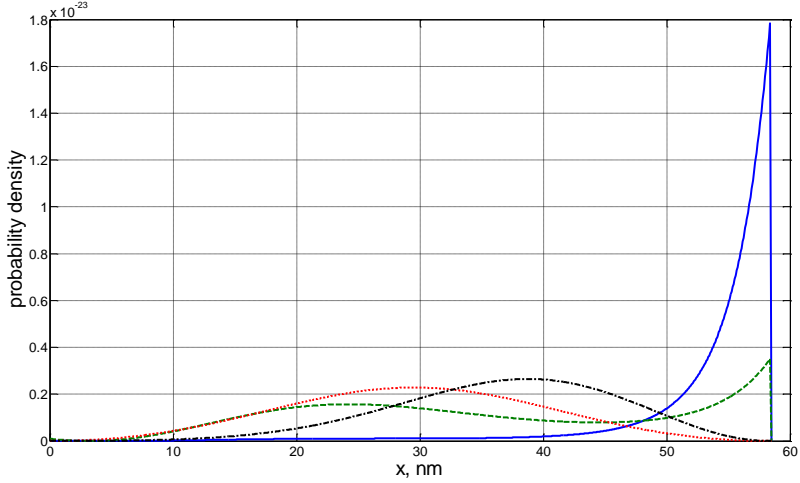

Fig. 17. Probability-density shapes in electron waveguide biased by electric field $E_0^{(x)}$. $B_0 = 2$ T , $U_0/E = 0.8667$ , and $a = 58.5$ nm . Solid blue line: $E_0^{(x)} = 0$ V/m ; green-dash line: $E_0^{(x)} = 3 \cdot 10^5$ V/m ; red-dot line: $E_0^{(x)} = 6 \cdot 10^5$ V/m ; black dot-dash line: $E_0^{(x)} = 9 \cdot 10^5$ V/m .

Then, this effect is similar to that found in Ref. 41 where a multilayer structure allows transforming the edge modes to non-localized ones by application of vertical-to-sandwich electric field. Both components (from Ref. 41) and our simple structure can be considered as the intermodal switches.

It is interesting to notice that varying the electric field, a basic intermediate shape is found (Fig. 17, red-dot line) which is similar to the one for waveguide biased only by constant confinement potential (Fig. 3, solid blue line). The electric field value providing this shape can be viewed as a critical one that corresponds in our case approximately $E_0^{(x)} = 6 \cdot 10^5$ V/m . Lower electric fields provide the localized modal shapes (Fig. 17, solid-blue and green-dot lines). Stronger electric field gives the waveforms of this basic shape distorted by the electric field (Fig. 17, black dot-dash line).

The electronic components based on the different mode signaling relate to the active-pause devices [35] because in their non-edge state, the signal (de-localized mode) is still at the output of a circuit. The best switches and signals are with the passive pause when in the off-state no energy appears at the output of a device. It decreases energy consumption and increasing the signal-to-noise ratio.

By simulation, it is found that a combination of proper chosen magnetic and electric fields allows to *switch electronically the edge-localized mode from its evanescent to the propagating condition and back in a simple way.* In this purpose, consider Fig. 3, where the main mode is becoming evanescent near $B_0 = 5.2$ T. Fig. 18 shows the increase of real propagation constant from its zero value with the applied electric field $E_0^{(x)}$ given for strong magnetic field $B_0 = 5.5$ T.

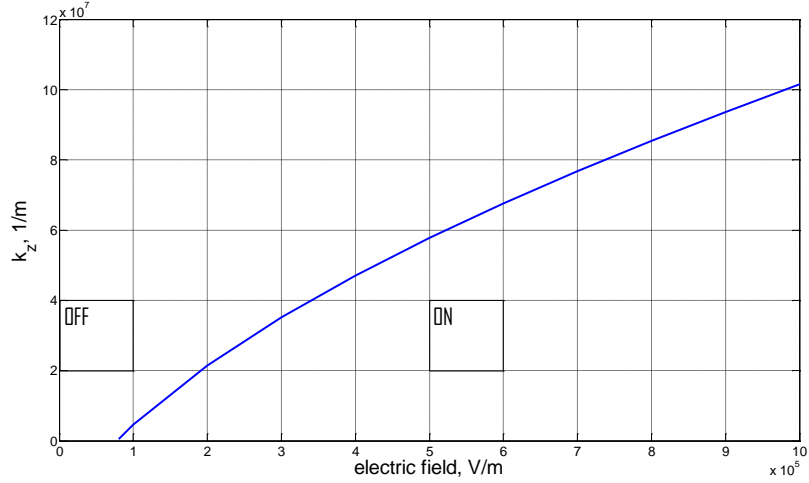

Fig. 18. Propagation constant $k_z$ of the edge-localized main mode versus electric field $E_0^{(x)}$. $U_0/E = 0.8667$, $B_0 = 5.5$ T, and $a = 58.5$ nm.

Thus, if $0 < E_0^{(x)} < 0.8 \cdot 10^5$ V/m, the main mode is not propagating, and the signal appears at the end of waveguide only due to tunneling effect that can be chosen negligible enough by lengthening of the waveguide or by increasing magnetic field.

For $E_0^{(x)} > 0.8 \cdot 10^5$ V/m, a large signal, which is transported by propagating the main mode, is at the end of waveguide length. For this wide range of electric field values $0.8 \cdot 10^5 < E_0^{(x)} < 10^6$ V/m and the chosen magnetic field $B_0 = 5.5$ T, the electric field does not harm the edge-localized shape of the main mode (Fig. 19), and this waveguide with its properly chosen parameters can work in switching regime (Table. 1).

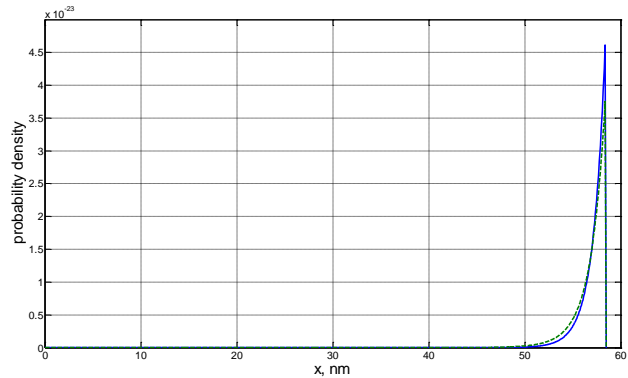

Fig. 19. Probability-density shapes in electron waveguide biased by electric field $E_0^{(x)}$. $B_0 = 5.5$ T, $U_0/E = 0.8667$, and $a = 58.5$ nm. Blue-solid line: $E_0^{(x)} = 2 \cdot 10^5$ V/m; green-dash line: $E_0^{(x)} = 10^6$ V/m.

Table. 1. Truth-table of proposed waveguide switch of edge modes

| $L$ | Control electric field $E_0^{(x)}$ level | Output logical state |
|---|---|---|
| 1 | $0 < E_0^{(x)} < 0.8 \cdot 10^5$ V/m | 0 |
| 2 | $0.8 \cdot 10^5$ V/m $< E_0^{(x)} < 10^6$ V/m | 1 |

To estimate the parameters of this switch connected to the source and load, it is necessary to calculate the device currents in the *on* and *off* regimes. For this purpose, similarly to [5,7], the Landauer-Buttiker formalism can be used. Further treatment of this effect can be found in [45] where the influence of randomness of confinement potentials and the electric and magnetic fields is considered on this switching mechanism.

## 8. Results and Conclusions

It has been shown how the applied transversal electric field can control the modal spectrum of the integer-Hall effect 2D waveguide, including the refining the waveguide's spectrum from the propagating non-localized waves and switching the rest traveling edge modes to their evanescent state and back. All these discovered effects are interesting in the development of new quantum coherent edge-mode logic components. The mentioned waveguides and the effects in them have been studied using the order reduction method in its Kron's circuit interpretation allowing semi-analytical modeling of waveguides arbitrary-biased by the confinement, magnetic and electric potentials. The effect of the potential randomness influencing the electronic modal spectrum control has been studied in details, which is interesting from the practical design point of view.